\documentstyle[aps,prl,epsf,floats]{revtex}

\draft

\begin{document}

\title{Oscillation effects on neutrino decoupling in the early universe}

\author{Steen Hannestad}

\address{NORDITA, Blegdamsvej 17, DK-2100 Copenhagen, Denmark}

\date{\today}

\maketitle

\begin{abstract}
In the early universe, 
neutrinos decouple from equilibrium with the electromagnetic plasma
at a temperature which is only slightly higher than the temperature
where electrons and positrons annihilate. Therefore neutrinos to
some extent share in the entropy transfer from $e^+e^-$ to other
species, and their final temperature is slightly higher than the
canonical value $T_\nu = (4/11)^{1/3} T_\gamma$. We study neutrino
decoupling in the early universe with effects of neutrino oscillations
included, and find that the change in neutrino energy density from
$e^+ e^-$ annihilations can be about 2-3\% higher if oscillation
are included. The primordial helium abundance can be changed by
as much as $1.5 \times 10^{-4}$ by neutrino oscillations.
\end{abstract}



\section{Introduction}

Freeze-out of particle species from thermal equilibrium is one of the
most important features of the early universe. One example is 
supersymmetric cold dark matter such as the neutralino. If these
particles were kept in thermal equilibrium throughout the history
of the universe their present day abundance would be suppressed
by a huge factor because of their large mass. However, because of
their weak interactions, the pair annihilation processes which keep
the particles in equilibrium stop being efficient as a temperature
of the order $T \sim m/20$ \cite{CDM}. At this time the particles decouple
from equilibrium and their abundance is afterwards only diluted
by cosmological expansion. Therefore such particles can be sufficiently
abundant today to make up the dark matter \cite{CDM}.

In the present paper we discuss the decoupling of neutrinos from
thermal equilibrium at a temperature of $\sim 1$ MeV.
In the canonical picture the electron neutrinos decouple at a temperature
of about 2 MeV, whereas the $\mu$ and $\tau$ neutrinos, which do not
have charged current interactions, decouple at roughly 4 MeV.
Shortly after this decoupling the temperature drops below the rest
mass of the electron and electrons and positrons pair annihilate 
and dump their entropy into photons. The neutrinos are not supposed
to share in this entropy transfer and therefore their 
temperature is lower than the photon temperature by a factor
$(4/11)^{1/3}$ \cite{kolbturner}.
It is also implicit in this treatment that oscillations between active
neutrino species have no effect because their distributions remain
identical.

However, because the temperature difference between neutrino decoupling
and electron-positron annihilation (at $T \sim m_e/3$) is only 
one order of magnitude the neutrinos will to some extent share in
the entropy transfer and the final temperature of the neutrinos will
be slightly different from the canonical value.

This problem has been treated many times in the past with various
methods \cite{Dicus:1982bz,Rana:1991xk,herrera,Dolgov:1992qg,Dodelson:1992km,Fields:1993zb,Hannestad:1995rs,Dolgov:1997mb,Dolgov:1999sf,gnedin,Esposito:2000hi,Steigman:2001px,Mangano:2001iu}. 
Early calculations made various approximations 
\cite{Dicus:1982bz,Rana:1991xk,herrera}, whereas
recent treatments have included solving the full momentum dependent
system of Boltzmann equations including all neutrino reactions
\cite{Dolgov:1992qg,Dodelson:1992km,Fields:1993zb,Hannestad:1995rs,Dolgov:1997mb,Dolgov:1999sf,gnedin,Esposito:2000hi}.

In all treatments so far, however, neutrinos have been assumed to be
non-mixed species (although the possibility that neutrino
heating could be changed by oscillations was mentioned
in Ref.~\cite{Langacker:1987jv}). 
The recent results from the Sudbury Neutrino Observatory (SNO) \cite{SNO}
 experiment have
confirmed that the Solar neutrino deficit is indeed explained by
active-active
neutrino oscillations, most likely $\nu_e - \nu_\mu$ oscillations
\cite{Barger:2001zs}.
However, the specific values of the mixing parameters have not
been conclusively measured. At present there are four different
solution regions in mixing parameter space. The best fit points
for these four solutions (from Ref.\ \cite{bahcall}) are listed in Table I.

\begin{table}
\caption{Best fit values of mixing parameters for solar neutrino
solutions, as well as their goodness of fit.}
\begin{center}
\begin{tabular}{lccc}
Solution & $\delta m^2/{\rm eV}^2$ & $\sin 2 \theta_0$ & goodness of fit \\ \tableline 
Large Mixing Angle (LMA) & $4.5 \times 10^{-5}$ & 0.91 & 59\% \\
Small Mixing Angle (SMA) & $4.5 \times 10^{-5}$ & $3.94 \times 10^{-2}$ & 19\% \\
Low & $1.0 \times 10^{-7}$ & 0.99 & 45\% \\
Vacuum (VAC) & $4.6 \times 10^{-10}$ & 0.91 & 42\%
\end{tabular}
\end{center}
\end{table}

Also, by far the best explanation for the atmospheric neutrino
problem is oscillations with maximal mixing
between $\nu_\mu$ and another neutrino,
most likely $\nu_\tau$, as suggested by the Super-Kamiokande
experiment \cite{superK}.
In the present paper we treat neutrino decoupling in the early
universe, taking into account neutrino oscillations. Unfortunately
the numerical complexity of the problem increases dramatically
if oscillations are introduced. We therefore treat the problem
in the so-called quantum rate equation approximation where the
momentum dependence of the problem is integrated out. We also refrain
from treating the full three-flavour oscillation problem and concentrate
on $\nu_e - \nu_\mu$ oscillations while assuming that $\nu_\tau$
is unmixed. Data on atmospheric neutrinos suggest that $\nu_\mu$
and $\nu_\tau$ are in fact maximally mixed, and our assumption is 
therefore not likely to hold. We will discuss this issue more
thoroughly in Section IV.

We find that neutrino heating is slightly more efficient if
oscillations are included. For non-oscillating neutrinos we find
an increase in neutrino energy density of $\Delta N_\nu \simeq 0.0467$,
where $\Delta N_\nu = \delta \rho/\rho_*$ and 
$\rho_* = \frac{7}{8}\left(\frac{4}{11}\right)^{4/3} \rho_\gamma$.
In the case of very efficient mixing
of $\nu_e$ and $\nu_\mu$ this is changed to $\Delta N_\nu \simeq 0.0479$,
a change of 2.5\%. We also find that oscillations can change the
primordial helium abundance. Neutrino heating induces a change of
$\Delta Y_P \simeq 1.2 \times 10^{-4}$ for non-oscillating
neutrinos. for maximal mixing this is changed to 
$\Delta Y_P \simeq 2.2 \times 10^{-4}$, i.e.\ a difference comparable
to the total magnitude of the effect.
In the limiting case of complete mixing of all neutrinos the result is
$\Delta N_\nu \simeq 0.0486$ and $\Delta Y_P \simeq 2.5 \times 10^{-4}$.

The next section contains a discussion of the essential rate
equations needed to treat neutrino decoupling, as well as the
equations needed to treat the cosmological expansion. 
Section III concerns the numerical details and results from solving
the equations, and Section IV contains a discussion and conclusion.
Finally, the Appendix contains various mathematical detail.


\section{Boltzmann equations}

The fundamental equation describing the evolution of all particle species
in the early universe is the Boltzmann equation, which for a non-mixed
species is \cite{bernstein}
\begin{equation}
\frac{df}{dt} = \sum_i C_i[f],
\end{equation}
where $C_i[f]$ are the collision terms. In a homogeneous and isotropic
system
$df/dt = \partial f / \partial t$. For the moment we do not
consider the expansion of the universe, but this will be discussed
at the end of the section.

However, for a treatment of oscillating neutrinos it is necessary to
replace the single particle distribution function with the density
matrix, $\rho$, for the mixed species. For neutrinos
the density matrix is a $3\times 3$ matrix, but in the present work
we limit ourselves to studying two-flavor oscillations, with the
third neutrino being non-mixed (in practice we study $\nu_e-\nu_\mu$ 
oscillations
with $\nu_\tau$ being non-mixed). The density matrix for neutrinos
can then be written as
\begin{equation}
\rho (p) = \frac{1}{2}[P_0(p)+{\bf P} (p) \cdot {\bf \sigma}],
\end{equation}
where $\sigma_i$ are the Pauli matrices. An equivalent expression
holds for anti-neutrinos.
Notice that this definition is slightly different from the notation
which is most often used 
$\rho(p)=\frac{1}{2} P_0(p)[1+{\bf P} (p) \cdot {\bf \sigma}]$. This notation
has the advantage that ${\bf P}$ describes the internal state of the
mixed neutrinos, whereas in our notation it does not. On the other hand
our notation significantly simplifies the equations because it is
linear (i.e. it does not contain a $P_0 {\bf P}$ term). 
The convention we use is the same as that used by McKellar and Thomson
\cite{McKellar:1994ja}
in their treatment of active-active oscillations in the 
early universe. The material in this section in follows their
treatment closely in several respects. 

The usual one-particle distribution functions are the diagonal
elements of the density matrix so that, 
$n_\alpha(p) = \frac{1}{2} [P_0(p)+P_z(p)]$ and
$n_\beta(p) = \frac{1}{2} [P_0(p)-P_z(p)]$.

The mixed neutrinos,
described by $(P_0,{\bf P})$ then evolve according to the equations
\begin{eqnarray}
\dot {\bf P} (p) & = & {\bf V} \times {\bf P} + [R_\alpha (p) - R_\beta(k)]
{\bf \hat z} - D(p) P_T(p) \\ && + \int dp' d(p,p') {\bf P}_T (p')
- {\bf C} (p) P_0 (p) +  \int dp' {\bf c}(p,p') P_0 (p') \label{eq:pol} \\
\dot P_0 (p) & = & R_\alpha(p) + R_\beta(p),
\end{eqnarray}
where ${\bf P}_T = P_x(p) {\bf \hat x} + P_y(p) {\bf \hat y}$ is the
transversal component of the polarization vector ${\bf P}$. The rate
terms are given by
\begin{eqnarray}
R_i (p) & = & \int dp' dk dk' \left[ \sum_j F_{ij} (pk | p'k')
[n_j (k') n_j (p') - n_i (k) n_i(p)] \right. \\
&& \left. - \frac{1}{2} \sum_l G_l (k'p' |
kp) {\bf P}_T (p) \cdot {\bf \bar P}_T^* (k) \right],
\end{eqnarray}
where $\sum_j$ is over all weakly interacting species and $\sum_l$
is over all other particles than the mixed neutrinos.
In all cases $\int dq$ is taken to mean integration over phase space
in the sense $\int dq = \int \frac{d^3 {\bf q}}{(2\pi)^3}$.

Here, the first term in the brackets is the usual Boltzmann equation
with exactly the same structure as for a non-mixed species. The second
term arises due to the possibility that mixed state neutrinos annihilate with
mixed state anti-neutrinos. This term therefore is only present for
active-active oscillations where both states are interacting.
The matrix element terms $F$ and $G$ are given by
\begin{eqnarray}
F_{ij}(pk | p'k') & = & 2 \pi N V^2(j(p),\bar j (k) |
i(p'),\bar i (k')) \delta (p+k - p' - k') \\
G_l (pk | p'k') & = & 2 \pi N V(\nu_\alpha (p'),
{\bar\nu}_\alpha (k') | l(k),\bar l (p)) \\
&& \times  V(\nu_\beta (p'),
{\bar\nu}_\beta (k') | l(k),\bar l (p))\delta (p+k - p' - k').
\end{eqnarray}
All the non-mixed species evolve according to the standard Boltzmann
equation
\begin{eqnarray}
\dot n_i (p) & = & \int dp' dk dk' \left[ \sum_j F_{ij} (pk | p'k')
[n_j (k') n_j (p') - n_i (k) n_i(p)] \right. \\ && \left. + G_i (kp |
k'p') {\bf P}_T (p) \cdot {\bf \bar P}_T^* (k) \right],
\end{eqnarray}
where again $\sum_j$ goes over all species.
As before there is a new term which arises from annihilation of mixed states.
The terms in Eq.~(\ref{eq:pol}) containing ${\bf c},{\bf C},D$ and $d$
are damping terms from elastic and inelastic scatterings which break
coherence of the oscillations. Details of how to calculate these terms
can be found in Ref.\ \cite{McKellar:1994ja}.

Finally, the ${\bf V} \times {\bf P}$ term is the usual oscillation
term which is responsible for the flavour oscillations.
The potential vector ${\bf V}$ can be written as
\begin{equation}
{\bf V} = 2 E_{\alpha \beta} {\bf \hat x} + (E_{\alpha \alpha} - 
E_{\beta \beta}) {\bf \hat z}.
\end{equation}
Here, 
\begin{eqnarray}
E_{\alpha \beta} & = &\frac{\delta m^2}{2 p} \sin 2 \theta_0 - 
V_{\alpha \beta} (p)\\
E_{\alpha \alpha} - E_{\beta \beta} & = & 
- \frac{\delta m^2}{2 p} \cos 2 \theta_0 + V_\alpha(p) - V_\beta(p),
\end{eqnarray}
where $\delta m^2 = m_2^2 - m_1^2$ and $\theta_0$ is the vacuum
mixing angle.
The matter potentials, $V$, arise from the neutrino interactions with
the medium \cite{Notzold:1988ik,Enqvist:1991ad}.

The above set of equations is quite complicated to solve, both because
of its non-linearity and because it is momentum dependent. However, 
it is simplified enormously by using the so-called quantum rate 
equations instead of the full quantum kinetic equations presented above.
In the quantum rate equations, the original quantum kinetic equations
are integrated over momentum space so that the momentum dependence
disappears. This integration can only be accomplished analytically 
if one assumes 
kinetic equilibrium for the neutrinos, and therefore involves an assumption.
We assume that the one-particle distribution functions for neutrinos
(the diagonal parts of the density matrix) are of the form $e^{-p/T}$.

The usual number densities of the mixed neutrinos are
then related to the integrated density matrix by
$n_\alpha = \frac{1}{2}[P_0+P_z]$ and $n_\beta = \frac{1}{2}[P_0-P_z]$.
However, this number density is not a dimensionless quantity. 
In order to make ${\bf P}$ and $P_0$ dimensionless we instead work
with the dimensionless quantities ${\bf P_*} \equiv {\bf P}/n_{\nu_0}$
and $P_{0,*} \equiv P_0/n_{\nu_0}$, where $n_{\nu_0}$ is the
number density of a decoupled neutrino species. As will be explained
at the end of the section this also has the advantage of making
the expansion of the universe simpler to treat.
For simplicity we will in the remainder of the paper refrain from
denoting ${\bf P_*}$ and $P_{0,*}$ with an $\ast$.

The quantum rate equations then take the form \cite{McKellar:1994ja}

\begin{eqnarray}
{\bf \dot P} & = & {\bf V} \times {\bf P} - D {\bf P}_T
- C {\bf \bar P}_T^* + [R_\alpha - R_\beta]
{\bf \hat z} \label{eq:rate1} \\
\dot P_0 & = & R_\alpha + R_\beta
\end{eqnarray}
where
\begin{eqnarray}
R_i & = & \sum_j F_{ij} [h_j n_j n_{\bar j} - n_{\nu_i} n_{\bar \nu_i}]
- \frac{1}{2} \sum_l G_l {\bf P}_T \cdot {\bf \bar P}_T^* \\
\dot n_i & = & \sum_j F_{ij} [h_j n_j n_{\bar j} - n_i n_{\bar i}]
+ G_i {\bf P}_T \cdot {\bf \bar P}_T^*.
\end{eqnarray}

$h_j=1$ for neutrinos and $h_j=\frac{1}{4}$ for electrons.
The new parameters $C,D,{\bf V}, F_{ij}$ and $G_i$ are then defined
as
\begin{eqnarray}
C & = & \frac{P_0}{{\bf \bar P}_T^*} \int dp 
\left[{\bf C} (p) n(p) - \int dp' {\bf c}(p,p') n(p') \right] \\
D & = & \int dp \left[D(p) n(p) - \int dp' d(p,p') n(p')\right] \\
{\bf V} & = & \int dp {\bf V}(p) n(p) \\
F_{ij} & = & \int dp dp' dk dk' F_{ij}(pk | p'k') n(p) n(k) \\
G_i & = & \int dp dp' dk dk' G_i(p'k' | pk) n(p) n(k) \label{eq:rate2}
\end{eqnarray}

Some details about how to perform the phase space integrals are given
in the Appendix.

In the present calculation we are interested only in a differential effect,
i.e.\ the difference between the actual neutrino density and the
density if neutrinos were completely decoupled. In that case, the
quantum statistics of the involved particles is not very important
\cite{Dodelson:1992km}. In the following we therefore approximate
all quantum statistics with Maxwell-Boltzmann statistics. This means 
that Bosons and Fermions have exactly the same behaviour.

In previous calculations of neutrino oscillations in the early universe
it has been assumed that the active neutrino species have the same
temperature as the electromagnetic plasma. However, that is not the 
case during electron-positron annihilation. We therefore have to operate
with different temperatures for all the different species.
Specifically we always assume a Maxwell-Boltzmann distribution with a 
temperature given by $T_{\rm eff,i} = T_0 (n_i/n_{\nu_0})^{1/3}$, where
$n_{\nu_0}$ and $T_0$ are the number density and temperature of a completely 
decoupled non-mixed neutrino species. $n_i$ is the actual density
of the given species. 
Electrons, positrons and photons are assumed to be in full thermal
equilibrium with the temperature $T_\gamma = T_e$.

All of the above equations have been derived assuming a non-expanding
universe.
If the universe expands then for the usual one particle Boltzmann
equation, the Liouville term $df/dt$ is changed from $\partial f / \partial t$
to $\partial f / \partial t - H p \partial f / \partial p$
\cite{bernstein}. Likewise, the 
left-hand side of the integrated Boltzmann equation is changed
from $\dot n$ to $\dot n + 3 H n$ \cite{bernstein}. 
However, the equations can be
made to look exactly like they do for the non-expanding case if
they are recast in comoving quantities. The momentum of a
particle redshifts with expansion as $p \propto R^{-1}$. For the
momentum dependent Boltzmann equation one then defines the 
comoving momentum as $p_* = p R$, a quantity that does not redshift.
The Liouville operator then becomes 
$df/dt = \partial f / \partial t - H p \partial f / \partial p =
\partial f(p_*) /\partial t$, i.e.\ it looks exactly like the 
non-expanding case.

For the integrated Boltzmann equation this also holds.
The usual procedure is to write the number densities in units
of entropy density as $n_* = n/s$. If entropy is conserved $n_*$
is not affected by cosmic expansion. However, in the present case
entropy is not conserved because full thermodynamic equilibrium
is not maintained. Instead one can rescale the number density
in units of a completely decoupled neutrino species, $n_{\nu_0} \propto
R^{-3}$. In this case the left hand side of the Boltzmann equation
reads $\dot n + 3 H n = \dot n_*$. If one recasts the
quantum rate equations in units of $n_{\nu_0}$ the cosmological
expansion does not appear anywhere, and one can readily use
Eqs.~(\ref{eq:rate1}-\ref{eq:rate2}) even in an expanding universe.

\subsection{Matter potentials}

The diagonal part of the matter potential comes both from interactions
with other species, and from self-interactions. 

The electron lepton number is known to be small ($L_e = L_p \sim 10^{-10}$)
because of charge neutrality. However the neutrino lepton numbers are
not well constrained at present. At present the strongest constraints
come from considering Big Bang nucleosynthesis (BBN)
and Cosmic Microwave Background Radiation (CMBR) arguments. 
From BBN considerations one finds an upper bound on relativistic
energy density (usually expressed as the effective number of neutrino
species $N_\nu \equiv \rho/\rho_{\nu_0}$)
\cite{Lisi:1999ng,Olive:2000ij}, 
and therefore also on neutrino
lepton numbers \cite{kang}. 
However, because electron neutrinos enter directly
into the weak interactions that regulate the neutron to proton ratio
an electron neutrino chemical potential cannot be directly translated
into an effective number of neutrino species \cite{kang}.
CMBR on the other hand is not sensitive to neutrino flavour but only
to energy density. Therefore the CMBR bound on the effective
number of neutrino species can be directly translated into a bound
on neutrino lepton numbers \cite{pastor,Hannestad:2000hc,Hannestad:2001hn}. 
Hansen {\it et al.} \cite{Hansen:2001hi} recently combined BBN
and CMBR to derive the tightest present constraint of
$-0.01 < L_{\nu_e} < 0.22$ and $|  L_{\nu_{\mu,\tau}} | < 2.6$
(see also \cite{Esposito:2000hh,Orito:2000zb,Esposito:2001sv,Kneller:2001cd,zentner,Cyburt:2001pq}.
This constraint is of course many orders of magnitude larger than
the known value of the electron lepton number.

For the sake of simplicity we
assume that all lepton numbers are of the same order as $L_e$. In that
case they can be ignored in the calculation. This simplifies the
numerical computations tremendously because the neutrino and anti-neutrino
sectors decouple (i.e. the equations describing them are identical).
However, we stress that this is not necessarily a good approximation, 
depending on the actual values of neutrino chemical potentials.

If one neglects all lepton numbers, the result for electron neutrinos is
\begin{equation}
V_{e} = - \frac{96 \sqrt{2} G_F}{\pi^2 m_{W}^2}
\left[T_{\nu_e} T_\gamma^4 + \frac{1}{4} (1-x) T_{\nu_e}^5 \right],
\end{equation}
where $x \equiv \sin^2 \theta_W \simeq 0.226$.
For the muon neutrino one finds a similar expression
\begin{equation}
V_{\mu} = - \frac{96 \sqrt{2} G_F}{\pi^2 m_{W}^2}
\left[\frac{1}{4} (1-x) T_{\nu_\mu}^5 \right]
\end{equation}
This is very close to what is found using Fermi-Dirac statistics for 
neutrinos. For FD statistics the front factor should be $\frac{49 \zeta(4)}
{45 \zeta(3)} \pi^2 \simeq 9.68$ \cite{Enqvist:1992qj}
instead of $96/\pi^2 \simeq 9.72$ for MB.
More details about the calculation of the matter potentials can be
found in the Appendix.

In addition to the diagonal part of the matter potential there is an
off-diagonal part due to neutrino-neutrino and neutrino-antineutrino
forward scattering \cite{pantaleone}. 
This term is proportional to $\int {\bf P} (p) dp
= {\bf P}$, so that in the quantum rate approximation the term
is identically zero. The off-diagonal term has the effect of synchronising
the oscillations of different neutrino modes. However, in the quantum
rate approximation the problem is reduced to following a single
``effective'' mode and therefore the oscillations of different modes
is in some sense already synchronized. 
In fact there is an additional off-diagonal term which is proportional
to $\rho_{\alpha \beta}$, but as will be explained in the Appendix this
term is always very small.

\subsection{Annihilation and damping terms}

The annihilation term for the process $i \bar i \leftrightarrow
j \bar j$ for particle $i$ can be written as
\begin{equation}
R_{ij} = F_{ij} [h_j n_j n_{\bar j} - n_{\nu_i} n_{\bar \nu_i}] = 
\frac{4 G_F^2}{\pi^3} T_{\nu_0}^5 B \left[\frac{T_j^8}{T_{\nu_0}^8}
- \frac{T_i^8}{T_{\nu_0}^8}\right]
\equiv F_0 B \left[\frac{T_j^8}{T_{\nu_0}^8}
- \frac{T_i^8}{T_{\nu_0}^8}\right],
\end{equation}
where $B$ depends on the specific process. Table II lists the value
of $B$ for different processes. Again, the front factor value 4 is
very close to what is found using FD statistics 
($\simeq 3.97$)
\cite{Enqvist:1992qj}.

\begin{table}
\caption{Values of $B$ for different annihilation processes.}
\begin{center}
\begin{tabular}{cc}
Process & $B$ \\ \tableline 
$\nu_e \bar \nu_e \leftrightarrow e^+e^-$ & $8 x^2 + 4 x + 1$ \\
$\nu_\mu \bar \nu_\mu \leftrightarrow e^+e^-$ & $8 x^2 - 4 x + 1$ \\
$\nu_\tau \bar \nu_\tau \leftrightarrow e^+e^-$ & $8 x^2 - 4 x + 1$ \\
$\nu_e \bar \nu_e \leftrightarrow \nu_\mu \bar \nu_\mu$ & 1 \\
$\nu_e \bar \nu_e \leftrightarrow \nu_\tau \bar \nu_\tau$ & 1 \\
$\nu_\mu \bar \nu_\mu \leftrightarrow \nu_\tau \bar \nu_\tau$ & 1
\end{tabular}
\end{center}
\end{table}

The same front factor can be used to 
calculate the damping coefficients $C$ and $D$. 
The calculation of these terms is discussed in Ref.\ \cite{McKellar:1994ja}
for the case of FD statistics.
Here we just state the result for MB statistics

\begin{eqnarray}
D & = &\frac{1}{2}F_0 \left[8 \frac{T_e^4 T_{\nu_e}^4}{T_{\nu_0}^8}
+ 8 \frac{T_e^4 T_{\nu_\mu}^4}{T_{\nu_0}^8} + 
(8 x^2 + 4 x + 1)\frac{T_e^4 T_{\nu_e}^4}{T_{\nu_0}^8} \right. \\ && \left. +
(8 x^2 - 4 x + 1)\frac{T_e^4 T_{\nu_\mu}^4}{T_{\nu_0}^8}+
2(\frac{T_{\nu_e}^8}{T_{\nu_0}^8}+\frac{T_{\nu_\mu}^8}{T_{\nu_0}^8}
+\frac{T_{\nu_e}^4 T_{\nu_\mu}^4}{T_{\nu_0}^8})\right]\\
C & = & 2 (16 x^2 + 4) F_0 \frac{T_{\nu_e}^4 T_{\nu_\mu}^4}{T_{\nu_0}^8}\\
G_e & = & (32 x^2 - 4) F_0 \frac{T_{\nu_e}^4 T_{\nu_\mu}^4}{T_{\nu_0}^8} \\
G_{\nu_\tau} & = & 4 F_0 \frac{T_{\nu_e}^4 T_{\nu_\mu}^4}{T_{\nu_0}^8}.
\label{eq:terms}
\end{eqnarray}
These expressions are very similar to those derived by
McKellar and Thomson \cite{McKellar:1994ja}
who used FD statistics but assumed identical
temperatures for all species.

\subsection{Time-temperature relationship}

Fundamentally, two equations are needed to fully describe the 
cosmological expansion with time
\cite{kolbturner}.
The system of photons and electrons/positrons can to an extremely good
approximation be assumed to be in full thermal equilibrium via
electromagnetic interactions, so that they can be described
by a common temperature, $T_\gamma$.
Therefore the two variables describing the cosmological expansion with
time can be taken to be the scale factor, $R$, and the photon
temperature, $T_\gamma$.

As the two independent equations we choose the Friedman equation 
\cite{kolbturner}
\begin{equation}
H^2 = \frac{8 \pi G \rho}{3}
\label{eq:fried1}
\end{equation}
and the equation of energy conservation
\cite{kolbturner}
\begin{equation}
d(\rho R^3) + P d(R^3) = 0
\label{eq:fried2}
\end{equation}
as our fundamental equations. 
These two equations can then be rewritten as equations for $\dot T_\gamma$
and $\dot R$.
Details of how to solve these equations
in the case of Maxwell-Boltzmann statistics can be found in
Ref.\ \cite{Dodelson:1992km}.


\section{Numerical details}

We have solved the quantum rate equations,
Eqs.~(\ref{eq:rate1}-\ref{eq:rate2}), together with the expansion equations
Eqs.~(\ref{eq:fried1}-\ref{eq:fried2}), for the case of $\nu_e-\nu_\mu$
oscillations. $\nu_\tau$ is in the present calculation assumed to be
non-mixed. As initial conditions we choose $T_\nu = T_\gamma = 15$ MeV,
a temperature well above the electron-positron annihilation temperature
$T_{\rm ann} \sim 0.3$ MeV. Furthermore we set $P_x = P_y = P_z = 0$
and $P_0 = 2$ (and identically for anti neutrinos). However, the 
outcome is not sensitive to initial conditions in the mixed neutrino
sector because at high temperatures ${\bf P}$ is quickly
driven to zero because of fast interactions, no matter what its
initial value is.
The system of equation is then straightforward to solve.

Fig.~1 shows the evolution of the quantity
\begin{equation}
\frac{\delta n}{n} \equiv \frac{n_\nu - n_{\nu_0}}{n_{\nu_0}}
\end{equation}
for $\nu_e,\nu_\mu$ and $\nu_\tau$,
for different values of the mixing angle, all calculated for the
specific value of the mass difference $\delta m^2 = 3 \times 10^{-5} 
\, {\rm eV}^2$. Again, $n_{\nu,0}$ is the number density of a standard,
decoupled neutrino species.
This figure also shows the
limiting case of non-mixed neutrinos ($\sin 2 \theta_0 = 0$).

It should be noted that the result of the non-mixed case is quite
close to the result found by elaborate momentum-dependent calculations.
The calculation by Dodelson and Turner \cite{Dodelson:1992km}
used the same approximation
as we have in the present work (i.e.\ zero electron mass and 
Boltzmann statistics for all particles), except that they solved
the full momentum dependent Boltzmann equation for the non-mixed case.
Their result was approximately $\delta \rho_{\nu_e}/\rho_{\nu_0}
\simeq 0.012$ and $\delta \rho_{\nu_\mu}/\rho_{\nu_0}
\simeq 0.005$, whereas our result for the non-mixed case is
$\delta \rho_{\nu_e}/\rho_{\nu_0}
\simeq 0.0124$ and $\delta \rho_{\nu_\mu}/\rho_{\nu_0}
\simeq 0.0057$.
Notice that $\delta \rho/\rho = (\rho_\nu-\rho_{\nu_0})/\rho_{\nu_0}$
is not equivalent to $\delta n/n$, the quantity shown in Fig. 1. 
However, since we assume thermal 
Maxwell-Boltzmann distributions for all particles the two 
quantities are simply related by $\delta \rho/\rho = \frac{4}{3} 
\delta n/n$.

This shows that although our calculation is quite
crude in the sense that it does not fully account for momentum dependence
it yields results which are fairly close to momentum dependent calculations,
at least for the case of non-mixed neutrinos. We do expect the same to
be true for the mixed case, although this remains to be verified.

Calculations which have used exact quantum statistics and electron
mass give slightly smaller neutrino heating. Hannestad and Madsen
found 0.0083 for $\nu_e$ and 0.0041 for $\nu_\mu$
\cite{Hannestad:1995rs}, whereas
Dolgov, Hansen and Semikoz found 0.009 and 0.004 \cite{Dolgov:1997mb,Dolgov:1999sf}. 
The most recent treatment by Gnedin and Gnedin found 0.0097 and
0.0062 \cite{gnedin}.

In Fig.~2 we show the evolution of $P_z = n_{\nu_e}-n_{\nu_\mu}$ for
different values of the mixing angle. 
It is clear that for this specific choice of $\delta m^2$ a very
large mixing angle is needed to achieve a noticeable effect.
$\delta m^2 = 1 \times 10^{-5} \, {\rm eV}^2$ is of the same
order of magnitude as the best fitting solutions for the 
LMA ($\delta m^2 = 4.5 \times 10^{-5} \, {\rm eV}^2$) and
SMA ($\delta m^2 = 4.7 \times 10^{-6} \, {\rm eV}^2$) solutions
to the solar neutrino problem. Therefore it is clear that
for the SMA solution where the best fit at present is
$\sin 2 \theta_0 \simeq 0.04$ there will be no noticeable effect
of neutrino oscillations on neutrino heating. On the other hand
for the LMA solution $\sin 2 \theta_0 \simeq 0.91$ which is
large to enough to give a significant change.

The evolution of $P_z$ with temperature
can be understood as follows. At high temperatures
oscillations are suppressed by the matter potential, and therefore 
$P_z$ evolves independently of the mixing angle. However at a 
certain temperature the matter mixing angle goes through
a maximum and oscillations become an important equilibration factor. 
The rate of
equilibration between the two species is to a rough approximation given by
$\Gamma_{\rm eq} 
\sim D \cos^2 2 \theta \sin^2 2 \theta$\cite{Raffelt:1996wa}. 
This should be compared
with the rate at which the abundances are driven apart by interactions with
the electron-positron plasma, $\Gamma_{\rm drive} \sim (R_e-R_\mu)$.
The matter mixing angle is given by the expression
\begin{equation}
\sin^2 2\theta = \sin^2 2\theta_0/(1- 2 f \cos
2 \theta_0 + f^2),
\end{equation}
where $f \equiv 6 T (V_e-V_\mu)/\delta m^2$.
This mixing angle goes through a maximum when
$f = \cos 2 \theta_0$ which also corresponds to a maximum
in the equilibration rate. This maximum occurs exactly
at the temperature
where the dip in $P_z$ is seen. Below this temperature the mixing
angle approaches the vacuum value and $\Gamma_{\rm eq}/\Gamma_{\rm drive}$
approaches a constant value, which to a reasonable approximation
is
\begin{equation}
\frac{\Gamma_{\rm eq}}{\Gamma_{\rm drive}} \to \frac{8}{16 x^2 + 2} 
\left(\frac{4}{11}\right)^{4/3} \sin^2 2\theta_0 \cos^2 2 \theta_0.
\end{equation}

This asymptotic value is always smaller than one so that for small
temperatures $P_z$ follows the same trend independently of the vacuum
mixing angle because the driving term is dominant.
However, as the vacuum mixing angle increases the equilibration around the
maximum of the mixing angle becomes more and more important. Therefore
the final value of $P_z$ decreases strongly with increasing 
vacuum mixing angle.

Oscillations in general become important once the matter potential
no longer dominates the vacuum oscillation term. This happens when
\begin{equation}
T_{\rm MeV} < \left(\frac{\delta m^2 \cos 2 \theta_0}{1.0 \times 10^{-7}
\, {\rm eV}^2}\right)^{1/6}
\end{equation}
For the large value of $\delta m^2$ used in Fig.~2 this temperature
is above the decoupling temperature for neutrinos. Therefore once
the oscillations become important they are quickly damped and no
oscillation pattern is seen at lower temperatures.

This is not the case for lower values of $\delta m^2$ where oscillations
only become important well after neutrino decoupling.
In Fig.~3 we show the evolution of $P_z$ for various values of
$\delta m^2$. For $\delta m^2 = 10^{-6} \, {\rm eV}^2$ the 
oscillations are completely damped away because neutrinos have not
decoupled before oscillations become important.
For $\delta m^2 = 10^{-8} \, {\rm eV}^2$ oscillations are apparent,
but still of very low amplitude. Notice that at low temperatures
the mean of the oscillating
curve still rises slowly due to neutrino heating by the electron-positron
annihilations.
Finally, for $\delta m^2 = 10^{-10} \, {\rm eV}^2$ neutrinos have
decoupled completely before oscillations become important. This means
that there is effectively no damping of neutrino oscillations 
once they commence.

Therefore, for low values of $\delta m^2$, $P_z$ does not approach 
a definite value for low temperatures, even though the total
neutrino number density, $P_0$, does.

Of course the oscillation of $P_z$ at low temperature is an artifact
of the quantum rate approximation. In the momentum dependent treatment
different modes oscillate with different frequencies and oscillations
decohere because of this effect. However, it should be noted that
if neutrino lepton numbers are significant, this picture can change
completely \cite{Pastor:2001iu}. 
In that case the off-diagonal elements in the neutrino
matter potential are large, and neutrino oscillations become
coherent.
As long as all lepton numbers can be neglected, as was assumed in
the present treatment, it will make better sense to use the
average of $P_z$ instead of the actual value because the decoherence
effect was not accounted for.

The increase in oscillation amplitude seen for decreasing 
$\delta m^2$ is also seen for active-sterile oscillations, for the
same reason \cite{Enqvist:1992qj}.

One non-standard feature in the active-active oscillations is the
appearance of the interference terms in the Boltzmann equation.
These terms could potentially be important and lead to a different
result for neutrino heating. However, it turns out that they are
always completely negligible compared with the usual collision
terms.
Fig.~4 shows the quantity $Q \equiv \frac{1}{2} 
\sum_i G_i {\bf P}_T \cdot {\bf \bar P}_T^*/
\sum_j F_{ij} [h_j n_j n_{\bar j} - 
n_{\nu_i} n_{\bar \nu_i}]$ for $i = \nu_e$,
for the specific case of 
$\delta m^2 = 3 \times 10^{-5} \, {\rm eV}^2$,
$\sin 2 \theta_0 = 0.5$. This clearly shows that
the non-standard terms from annihilation of mixed states is tiny compared
with the standard Boltzmann terms and that they can be safely ignored
in numerical treatments.


\section{Discussion}

We have solved the Boltzmann equations governing the evolution of
neutrinos around the time of their decoupling from equilibrium, including
effects due to mixing of $\nu_e$ and $\nu_\mu$. We always assumed that
the tau neutrino is unmixed.

As could be expected oscillations have the effect of bringing the
$\nu_e$ and $\nu_\mu$ abundances closer together. Strong oscillations
also have an effect on the total neutrino energy density
after electron-positron annihilation. 
The reason is that without
oscillations, most of the heating is to the $\nu_e$ sector because
electron neutrinos have charged current interactions. Oscillations
drain away electron neutrinos into muon neutrinos, and therefore
the back-reaction $\nu_e \bar \nu_e \to e^+ e^-$ decreases in efficiency.
The end result is a slightly larger neutrino energy density.
As is customary we parametrise the neutrino energy density in units
of the energy density of a decoupled massless neutrino,
$N_{\nu} = \rho_\nu/\rho_{\nu_0}$ (so that in the
absence of neutrino heating $N_\nu = 3$).
However, it is also necessary to account for the slightly lower
photon temperature if neutrino heating is accounted for, because
all quantities should be measured relative to the actual photon temperature.
We therefore use the definition
\begin{equation}
N_{\nu} = \frac{\rho_\nu}{\rho_{\nu_0}} \frac{\rho_{\gamma_0}}
{\rho_\gamma}
\end{equation}

In Fig.~5 we plot the $N_\nu$ at low temperature. 
From this figure it can be seen that neutrino oscillations can
change the effective number of neutrino species by about
$1.2 \times 10^{-3}$. The effective number of neutrino species without
oscillations is $N_\nu = 3.0467$, and in the limit of large mass
difference and mixing angle it approaches $N_\nu = 3.0479$.
This value is somewhat higher than what is found in the
more thorough calculations using FD statistics and the full
momentum dependent Boltzmann equation 
($N_\nu = 3.028$ \cite{Hannestad:1995rs}, $N_\nu = 3.034$
\cite{Dolgov:1997mb,Dolgov:1999sf}, $N_\nu = 3.032$ \cite{gnedin})
simply because of the larger neutrino heating when MB
statistics is used (our value fits very well with that of
Ref.\ \cite{Dodelson:1992km}, who found $N_\nu = 3.046$ using
MB statistics).

For BBN calculations, there is an additional effect which must be
considered. The electron neutrinos have a different effect on
BBN than muon or tau neutrinos because they enter directly in
the $\beta$-reaction which regulate the neutron to proton ratio.
An increased number of electron neutrinos and anti-neutrinos 
have the effect of increasing the $n-p$ conversion efficiency.
This in turn leads to a lower neutron to proton ratio at helium formation
and in turn a lower helium abundance. This effect works in the opposite
direction of a simple increase in $N_\nu$.
Furthermore, because of energy conservation the photon temperature
is slightly lower if neutrino heating is included because photon
heating by $e^+ e^-$ annihilation is slightly smaller. This also
has the effect of lowering the $n-p$ conversion rate because
there are slightly fewer electrons and positrons.

To get a feeling for how neutrino oscillations change the helium
production we have modified the Kawano nucleosynthesis code 
\cite{Kawano:1992ua} to
take into account neutrino heating. We have then performed the
numerical calculation for oscillation parameters corresponding to 
the best fit for the LMA solar neutrino solution
($\sin 2 \theta_0 = 0.908$ and $\delta m^2/{\rm eV}^2 = 4.5 \times 10^{-5}$).
The result is shown in Table III, together with our result for the
case of no oscillations and the limiting case of large $\delta m^2$
and $\sin 2 \theta_0$.

\begin{table}
\caption{Change in helium abundance due to neutrino heating, for
the case of $\eta_{10} = 5 \times 10^{-10}$.}
\begin{center}
\begin{tabular}{lcccccc}
Solution & $\delta m^2/{\rm eV}^2$ 
& $\sin 2 \theta_0$ & $\delta \rho_{\nu_e}/\rho_{\nu_0}$
& $\delta \rho_{\nu_\mu}/\rho_{\nu_0}$ &
$\delta \rho_{\nu_\tau}/\rho_{\nu_0}$ & $\Delta Y_P$ \\ \tableline 
LMA & $4.5 \times 10^{-5}$ & 0.908 & 0.0097 & 0.0087 & 0.0057 & 
$2.0 \times 10^{-4}$ \\
No osc. & - & - & 0.0124 & 0.0057 & 0.0057 & $1.2 \times 10^{-4}$ \\
Maximal mixing & - & - & 0.0093 & 0.0093 & 0.0058 & $2.2 \times 10^{-4}$
\end{tabular}
\end{center}
\end{table}

This shows that more helium is produced for the oscillating
case. The main reason for this is the lower density of electron
neutrinos in the case of mixing, but the increase in the effective number of
neutrino species also leads to an increase in helium production.
Interestingly, the change in helium abundance due to oscillations
is of the same order of magnitude as the total effect. The reason is
that in the non-oscillating case the effect on helium is very small
because of cancellations. The increase in electron neutrino temperature
is roughly compensated by the increase in the effective number of
neutrinos, as well as the decrease in photon temperature.
Oscillations destroy this accidental cancellation and therefore have
a large effect.
Our finding for the non-oscillating case fits well with other calculations.
Using the same approximations as in the present paper Fields, Dodelson
and Turner \cite{Fields:1993zb} found $\Delta Y_P = 1.1 \times 10^{-4}$. 
More sophisticated methods find similar values in the range 
$\Delta Y_P = 1.1-1.5 \times 10^{-4}$

In terms of energy density
the changes to neutrino heating by neutrino oscillations are quite
small. If indeed the large mixing angle solution turns out to be 
correct then the effective number of neutrino species is changed
by roughly $8 \times 10^{-4}$ compared to the non-oscillating case. 
From CMBR the present bound on the effective number of neutrino
species is $N_\nu < 13$ \cite{Hannestad:2000hc,Hannestad:2001hn}, 
i.e.\ more than two orders of magnitude larger
than the effect induced by neutrino heating, and about $10^4$ times
bigger than the change induced by oscillations.
With precision data from upcoming satellite experiments such
as the Planck Surveyor it could be possible to measure 
$\Delta N_\nu \simeq 0.04$ \cite{Lopez:1999aq}
which is comparable to the effect from
neutrino heating. Even so the small difference induced by oscillations
will likely remain undetectable.

For BBN the change due to neutrino heating is of the order $10^{-4}$.
At present the observational uncertainty on the primordial helium
abundance is about $\sigma (Y_P) \sim 0.005$
\cite{Olive:2000ij}, which is about 50
times larger than the change. Here, however, the change due to neutrino
heating is more significant, comparable in magnitude to the
total neutrino heating effect. It is perhaps conceivable that the difference
in neutrino heating from including oscillations could be detected.

Finally, we stress that the present treatment is by no means definitive.
A proper treatment of momentum dependence is missing, as is the 
inclusion of three-neutrino oscillations. Inclusion of momentum
dependence is not likely to have a big effect for the relatively
large mass difference and mixing angle characterising the LMA solution,
but may be important for the vacuum solution. If $\nu_\mu$ and
$\nu_\tau$ are maximally mixed, as is indicated by atmospheric
neutrino measurements, then they will behave effectively as one
species during neutrino decoupling. This should have the effect
of lowering the electron neutrino temperature more than for 
two-neutrino oscillations, while increasing the overall effective
number of neutrino species slightly. Altogether this amounts
to the same effect as for two-neutrino oscillations, but slightly
larger.

It is simple to calculate the extreme upper limit on the oscillation
effect by considering a complete coupling of $\nu_\mu$ and $\nu_\tau$.
In Table IV we show results of the same calculation as in Table III,
but now assuming an infinitely tight coupling between muon and tau
neutrinos. This is likely to be a very good approximation to the
true state of affairs, because the preferred values of the mixing
parameters for $\nu_\mu-\nu_\tau$ mixing is at present \cite{superK}
($\delta m^2 \simeq 2 \times 10^{-3} \, {\rm eV}^2$, $\sin 2 \theta_0 
\simeq 1$). This means that oscillations become important already
when $T \simeq 5$ MeV, long before neutrinos decouple and also
long before neutrino heating commences. 
Therefore
in this case, $\nu_\mu$ and $\nu_\tau$ should be treated as 
having effectively the same temperature.
Note that this is only a good approximation when $\theta_{13}$ is small
so that there is little direct mixing of $\nu_e$ and $\nu_\tau$. 
Observations indeed indicate that this is the case.

\begin{table}
\caption{Change in helium abundance due to neutrino heating, for
the case of $\eta_{10} = 5 \times 10^{-10}$. Results are from
assuming infinitely tight coupling between $\nu_\mu$ and $\nu_\tau$.}
\begin{center}
\begin{tabular}{lcccccc}
Solution & $\delta m^2/{\rm eV}^2$ 
& $\sin 2 \theta_0$ & $\delta \rho_{\nu_e}/\rho_{\nu_0}$
& $\delta \rho_{\nu_\mu}/\rho_{\nu_0}$ &
$\delta \rho_{\nu_\tau}/\rho_{\nu_0}$ & $\Delta Y_P$ \\ \tableline 
LMA & $4.5 \times 10^{-5}$ & 0.908 & 0.0089 & 0.0077 & 0.0077 & 
$2.3 \times 10^{-4}$ \\
No osc. & - & - & 0.0124 & 0.0057 & 0.0057 & $1.2 \times 10^{-4}$ \\
Maximal mixing & - & - & 0.0082 & 0.0082 & 0.0082 & $2.5 \times 10^{-4}$
\end{tabular}
\end{center}
\end{table}

Maximal coupling between the muon and tau neutrinos therefore lead to 
a slightly larger increase in energy density due to neutrino heating.
In the case of maximal $\nu_e - \nu_\mu$ coupling it increase $N_\nu$
from 3.0479 to 3.0484, and for the LMA solution from 3.0476 to
3.0478. Helium production is also slightly increased, by about 
$0.2 \times 10^{-4}$. Although we have not performed a full three-neutrino
oscillation calculation, this estimate should be fairly close to the
true value because $\nu_\mu$ and $\nu_\tau$ are most likely maximally
mixed with a large mass difference \cite{superK}.

\acknowledgements
I wish to thank Dmitri Semikoz, Sergio Pastor, and Georg Raffelt for
valuable comments.


\section{Appendix}

\subsection{Matter potentials}

The diagonal 
matter potentials arise from neutrino loop interactions with the 
background medium. The magnitude of the potentials were first
calculated by N{\"o}tzold and Raffelt \cite{Notzold:1988ik}
(see also \cite{Enqvist:1991ad}). 
In the absence of lepton
numbers, there is no contribution due to interactions with neutrinos
of different flavour. Neutrinos of the same flavour yield the
contribution
\begin{equation}
V = \frac{16 \sqrt{2} G_F p}{3 m_Z^2} \langle E_\nu \rangle N_\nu.
\end{equation}
For electron neutrinos there is an additional contribution from
interactions with the background electrons and positrons.
If the electron mass is neglected the contribution is
\begin{equation}
V = \frac{16 \sqrt{2} G_F p}{3 m_W^2} \langle E_e \rangle N_e.
\end{equation}
Neglecting the electron mass in the matter potentials only leads
to very small errors and is consistent with neglecting it in
the interaction matrix elements.
In the quantum rate approximation one should make the
replacement $p \to \langle E_\nu \rangle = 3 T_\nu$. Then using
$N_e = 2 T_\gamma^3/\pi^2$ and $N_\nu = T_\nu^3/\pi^2$, one
finds 
\begin{equation}
V_{e} = - \frac{96 \sqrt{2} G_F}{3 \pi^2 m_{W}^2}
\left[T_{\nu_e} T_\gamma^4 + \frac{1}{4} (1-x) T_{\nu_e}^5 \right]
\end{equation}
for the electron neutrinos.
For the muon neutrino there is no contribution from electrons, and
one finds the result
\begin{equation}
V_{\mu} = - \frac{96 \sqrt{2} G_F}{3 \pi^2 m_{W}^2}
\left[\frac{1}{4} (1-x) T_{\nu_\mu}^5 \right]
\end{equation}

As was mentioned in Section II.a there is an additional off-diagonal
term which is of the form \cite{McKellar:1994ja}

\begin{equation}
V = \frac{8 \sqrt{2} G_F p}{3 m_Z^2} 
\langle E_\nu \rangle \rho_{\alpha \beta}.
\end{equation}

However, because the off-diagonal elements of $\rho_{\alpha \beta}$ are
always very small until long after neutrino freeze-out this contribution
is negligible.

\subsection{Phase-space integrals}

In this section we discuss how to perform the phase-space integrals
needed to calculate the terms in Eqs.~(\ref{eq:terms}). A full derivation
of all the terms would be too lengthy, but as a representative example
we show the calculation of the contribution to $R_e$ by the
process $\nu_e \bar \nu_e \leftrightarrow e^+ e^-$.
\begin{eqnarray}
R_e (\nu_e \bar \nu_e \leftrightarrow e^+ e^-) & = &
\frac{1}{n_{\nu_0}}
\int dp dp' dk dk' (2\pi)^4 V^2(pk | p'k') \delta(p+k-p'-k')\nonumber\\
&& \,\,\, [f_e (p') f_e(k') - f_{\nu_e} (p)f_{\nu_e} (k)]
\end{eqnarray} 
Here, $dp \equiv \frac{d^3 p}{2 p_0 (2\pi)^3}$, $f_e(p') = e^{-p'/T_e}$,
and $f_{\nu_e} (p) = e^{-p/T_{\nu_e}}$. All quantities are normalized
to the density of a single decoupled neutrino species, which explains
the $\frac{1}{n_{\nu_0}}$ in front.

The squared matrix element is
\begin{equation}
V^2(pk | p'k') = 32 G_F^2 [(2x+1)^2 (p \cdot k')(k \cdot p')+
(2x)^2 (p \cdot p')(k \cdot k')].
\end{equation}
Because of the Boltzmann statistics, $f_e (p') f_e(k') = f_e(p) f_e(k)$.
Using this, the $dp',dk'$ integrals can be performed using Lenard's
formula
\begin{equation}
\int \frac{d^3 p'}{2 p_0'}\frac{d^3 k'}{2 k'_0} \delta(P-p'-k')
p'^\mu k'^\nu = \frac{\pi}{24} (2P^\mu P^\nu + g^{\mu \nu} P^2)
\end{equation}
The result is
\begin{eqnarray}
R_e (\nu_e \bar \nu_e \leftrightarrow e^+ e^-) & = & \frac{1}{n_{\nu_0}} \frac{16}{3}
\frac{\pi G_F^2}{(2\pi)^8} (8 x^2 + 4 x + 1) \\ && \,\, \times \int 
\frac{d^3 p}{2 p_0}
\frac{d^3 k}{2 k_0} 
[e^{-p/T_e} e^{-k/T_e} - e^{-p/T_{\nu_e}} e^{-k/T_{\nu_e}}] (p \cdot k)^2
\nonumber 
\end{eqnarray}
We then use that $(p \cdot k)^2 = p^2 k^2 (1-\cos \theta)^2$, where $\theta$
is the angle between the direction of $p$ and $k$.
After performing the integrals over $d^3 p$ and $d^3 k$ the result is then
\begin{equation}
R_e (\nu_e \bar \nu_e \leftrightarrow e^+ e^-) = \frac{1}{n_{\nu_0}} \frac{4 G_F^2}{\pi^5} 
(8 x^2 + 4 x + 1) [T_e^8 - T_{\nu_e}^8]
\end{equation}

Using $n_{\nu_0} = T_{\nu_0}^3/\pi^2$ 
the contribution to the Boltzmann collision integral is
\begin{equation}
R_e (\nu_e \bar \nu_e \leftrightarrow e^+ e^-) = \frac{4 G_F^2}{\pi^3} T_{\nu_0}^5 
(8 x^2 + 4 x + 1) \left[\frac{T_e^8}{T_{\nu_0}^8} - 
\frac{T_{\nu_e}^8}{T_{\nu_0}^8}\right]
\end{equation}

This result is almost identical to what is found using FD statistics
(3.97 in the front factor instead of 4).

\begin{figure}[h]
\begin{center}
\epsfysize=12truecm\epsfbox{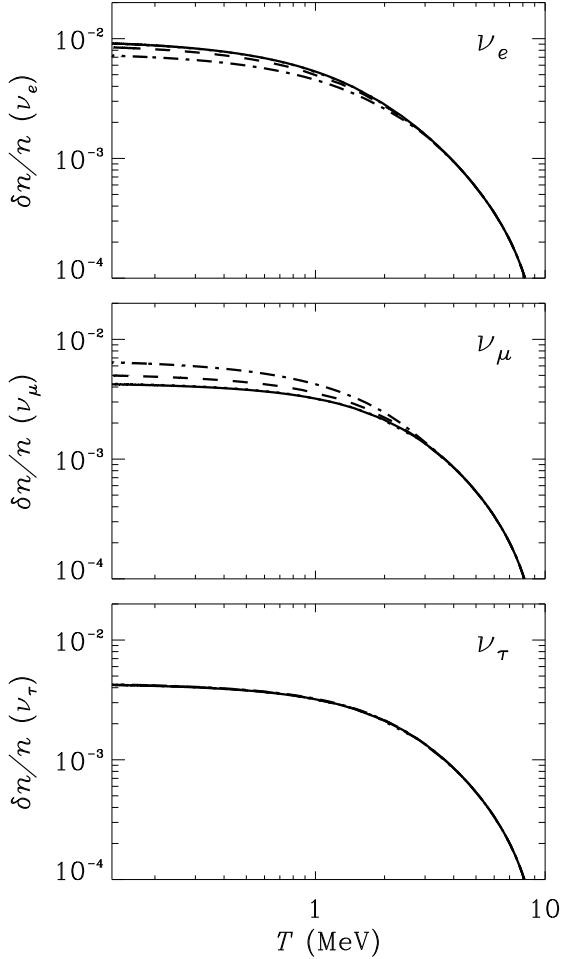}
\vspace{0truecm}
\end{center}
\vspace*{1cm}
\caption{The evolution of $\delta n/n$ for electron, muon and tau
neutrino for $\delta m^2 = 3 \times 10^{-5} {\rm eV}^2$, plotted
for different values of the vacuum mixing angle. The full line
is for $\sin 2 \theta_0 = 0$, the dotted for $\sin 2 \theta_0 = 0.1$,
the dashed for $\sin 2 \theta_0 = 0.5$ and the dot-dashed for
$\sin 2 \theta_0 = 0.9$.}
\label{fig1}
\end{figure}

\begin{figure}[h]
\begin{center}
\epsfysize=7truecm\epsfbox{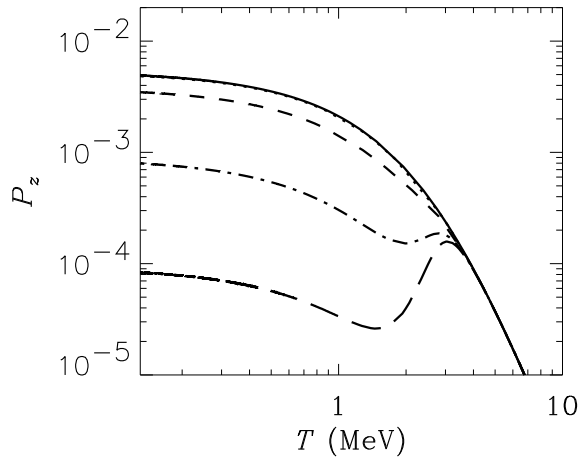}
\vspace{0truecm}
\end{center}
\caption{The temperature evolution of $P_z \equiv n_{\nu_e}-n_{\nu_\mu}$
for $\delta m^2 = 1.0 \times 10^{-5} {\rm eV}^2$ for various different
values of $\sin 2 \theta_0$. The full line is for 
$\sin 2 \theta_0 = 0$, the dotted for $\sin 2 \theta_0 = 0.1$, the
dashed for $\sin 2 \theta_0 = 0.5$, the dot-dashed for 
$\sin 2 \theta_0 = 0.9$, and the long-dashed for $\sin 2 \theta_0 = 0.99$.}
\label{fig2}
\end{figure}

\begin{figure}[h]
\begin{center}
\epsfysize=12truecm\epsfbox{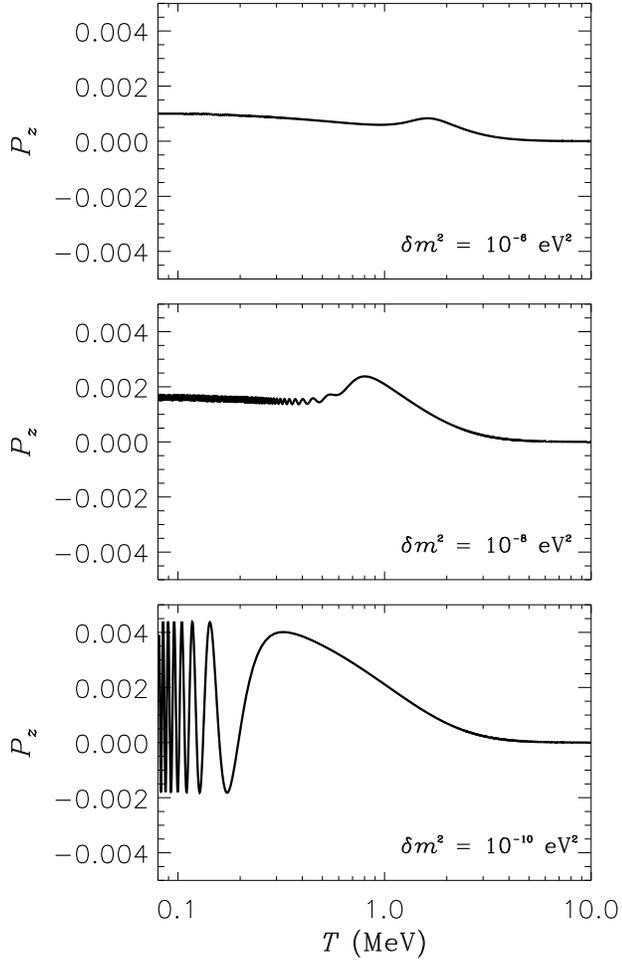}
\vspace{0truecm}
\end{center}
\vspace*{1cm}
\caption{The temperature evolution of $P_z \equiv n_{\nu_e}-n_{\nu_\mu}$
for $\sin 2 \theta_0 = 0.9$ for various different values of
$\delta m^2$. The upper panel is for 
$\delta m^2 = 1.0 \times 10^{-6} {\rm eV}^2$, the middle for
$\delta m^2 = 1.0 \times 10^{-8} {\rm eV}^2$ and the lower for
$\delta m^2 = 1.0 \times 10^{-10} {\rm eV}^2$.}
\label{fig3}
\end{figure}

\begin{figure}[h]
\begin{center}
\epsfysize=7truecm\epsfbox{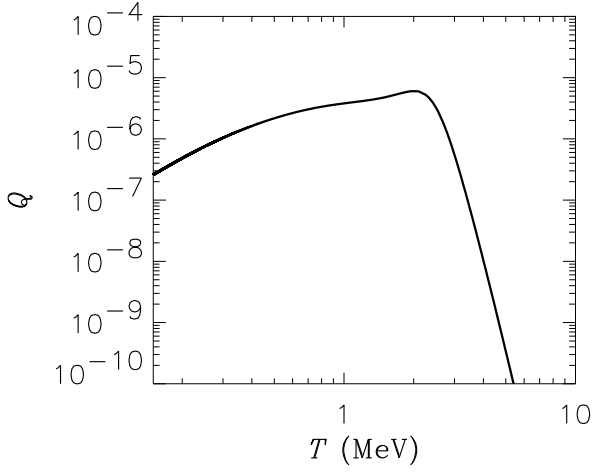}
\vspace{0truecm}
\end{center}
\caption{The evolution of the parameter $Q \equiv \frac{1}{2} 
\sum_i G_i {\bf P}_T \cdot {\bf \bar P}_T^*/
\sum_j F_{ij} [h_j n_j n_{\bar j} - 
n_{\nu_i} n_{\bar \nu_i}]$ for $i=\nu_e$, for the specific case of 
$\delta m^2 = 3 \times 10^{-5} \, {\rm eV}^2$,
$\sin 2 \theta_0 = 0.5$.}
\label{fig4}
\end{figure}

\begin{figure}[h]
\begin{center}
\epsfysize=7truecm\epsfbox{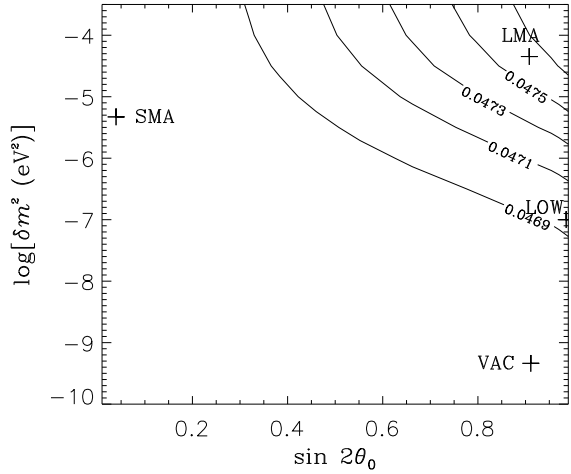}
\vspace{0truecm}
\end{center}
\caption{The total change in effective number of neutrino species
($\Delta N_\nu = N_\nu - 3$) as a function of $\delta m^2$ and
$\sin 2 \theta_0$. Shown are also the best fit values for the
possible solar neutrino solutions \protect\cite{bahcall}.}
\label{fig5}
\end{figure}

\end{document}